\newif\ifJOURNAL
\JOURNALfalse
\newif\ifarXiv
\arXivfalse
\newif\ifWP
\WPfalse
\newif\ifFULL
\FULLfalse
\newif\ifSemiFULL
\SemiFULLfalse
\newif\ifLATIN
\LATINfalse

\arXivtrue

\ifFULL\SemiFULLtrue\fi

\ifarXiv\LATINtrue\fi	

\newif\ifnotJOURNAL	
\notJOURNALtrue
\ifJOURNAL\notJOURNALfalse\fi

\newif\ifnotarXiv	
\notarXivtrue
\ifarXiv\notarXivfalse\fi

\newif\ifTR		
\TRfalse
\ifarXiv\TRtrue\fi
\ifWP\TRtrue\fi
\newif\ifnotTR
\notTRtrue
\ifarXiv\notTRfalse\fi
\ifWP\notTRfalse\fi

\newif\ifnotFULL	
\notFULLtrue
\ifFULL\notFULLfalse\fi

\newif\ifnotLATIN	
\notLATINtrue
\ifLATIN\notLATINfalse\fi

\ifnotLATIN
  \newcommand{\Aleksandrov}{aleksandrov:1977}
  \newcommand{\Takeuchi}{takeuchi:2004}
\fi
\ifLATIN
  \newcommand{\Aleksandrov}{aleksandrov:1977latin}
  \newcommand{\Takeuchi}{takeuchi:2004latin}
\fi

\ifJOURNAL
  \documentclass{elsart}

  \usepackage{amsmath,amsfonts,amssymb,latexsym}
  \newcommand{\Extra}[1]{}
\fi

\ifarXiv
\documentclass{article}
\usepackage{amsmath,amsthm,amsfonts,amssymb,latexsym}
\newcommand{\Extra}[1]{}
\fi

\ifWP
\documentclass[toc]{gtarticle}
\usepackage{amsmath,amsthm,amsfonts,amssymb,latexsym,epsfig}
\renewcommand{\Extra}[1]{}
\fi

\ifFULL
\usepackage{color}
\renewcommand{\Extra}[1]{\blue{#1}}

\newcommand{\blue}[1]{\textcolor{blue}{#1}}
\newcommand{\bluebegin}{\begingroup\color{blue}}
\newcommand{\blueend}{\endgroup}

\fi

\allowdisplaybreaks[4]

\newcommand{\Vladimir}{Vladimir}
\newcommand{\DOT}{.}

\ifnotLATIN
  \input{OT2enc.def}
  
  \usepackage{CJK}
\fi

\newcommand{\st}{\mathrel{|}}		

\newcommand{\dd}{\mathrm{d}}		

\newcommand{\K}{\mathcal{K}}		

\newcommand{\FFF}{\mathcal{F}}		
\newcommand{\LLL}{\mathcal{L}}		

\DeclareMathOperator{\III}{\mathbb{I}}		

\newcommand{\bbbp}{\mathbb{P}}		
\DeclareMathOperator{\Prob}{\bbbp}
\DeclareMathOperator{\UpProb}{\overline{\bbbp}}		

\newcommand{\bbbe}{\mathbb{E}}		
\DeclareMathOperator{\Expect}{\bbbe}

\newcommand{\bbbr}{\mathbb{R}}		

\newlength{\IndentI}
\newlength{\IndentII}
\newlength{\IndentIII}
\newlength{\IndentIV}
\newlength{\WidthI}
\newlength{\WidthII}
\newlength{\WidthIII}
\newlength{\WidthIV}
\theoremstyle{plain}
\newtheorem{theorem}{Theorem}
\newtheorem{proposition}{Proposition}
\newtheorem{corollary}{Corollary}
\newtheorem{lemma}{Lemma}

\theoremstyle{definition}
\newtheorem{remark}{Remark}

\ifarXiv
\title{Continuous-time trading and\\emergence of randomness}
\author{Vladimir Vovk\\
\texttt{vovk{\rm@}cs.rhul.ac.uk}\\
\texttt{http://vovk.net}}
\fi

\ifWP
\title{Continuous-time trading and\\emergence of randomness}
\author{Vladimir Vovk}

\twodatestrue

\fi

\begin{document}
\ifJOURNAL
\begin{frontmatter}



\title{Continuous-time trading and\\emergence of randomness}


\author{Vladimir Vovk}

\address{Computer Learning Research Centre,
  Department of Computer Science,\\
  Royal Holloway, University of London,
  Egham, Surrey TW20 0EX, UK}
\ead{vovk@cs.rhul.ac.uk}
\ead[url]{http://www.vovk.net}
\fi

\ifnotJOURNAL
  \maketitle
\fi

\begin{abstract}
  A new definition of events of game-theoretic probability zero
  in continuous time
  is proposed and used to prove results
  suggesting that trading in financial markets
  results in the emergence of properties
  usually associated with randomness.
  This paper concentrates on ``qualitative'' results,
  stated in terms of order (or order topology)
  rather than in terms of the precise values taken by the price processes
  (assumed continuous).
\end{abstract}

\ifJOURNAL
\end{frontmatter}
\newpage
\fi

\section{Introduction}
\label{sec:introduction}

This paper proposes (in Section \ref{sec:definitions})
a new definition of continuous-time events of zero game-theoretic probability.
The applications (Sections \ref{sec:level} and \ref{sec:increase})
are to an idealized securities market,
with a security price modelled as a continuous process.
We show that the price path will, almost surely,
satisfy various properties usually associated with randomness.
The phrase ``almost surely'' refers to the fact
that a speculator can become arbitrarily rich risking only 1 monetary unit
if the price path does not behave this way;
therefore, if we believe that the market is to some degree efficient,
we expect that those properties will be satisfied.

We consider some of the standard properties
of typical sample paths of Brownian motion
usually found in probability textbooks
(such as \cite{karatzas/shreve:1991}, Section 2.9).
This paper is inspired by \cite{takeuchi/etal:2007},
which in turn develops some ideas in \cite{GTP5};
both those papers attempt to formalize the ``$\sqrt{\dd t}$ effect''
(the fact that a typical change in the value of a non-degenerate diffusion process
over time period $\dd t$ has order of magnitude $\sqrt{\dd t}$).
We, however, concentrate on those properties
that depend only on the ordering of the security prices
at different times,
rather than on the actual values of the prices.
Among such properties are, for example,
the absence of isolated zeroes of the price path
and the absence of points of strict increase or decrease.
The difference of the game-theoretic treatment
from the standard results
is that we do not assume \emph{a priori} any stochastic picture;
we start instead from a simple trading protocol
without making any probabilistic assumptions.

This paper is part of the recent revival of interest in game-theoretic probability
(whose idea goes back to Ville \cite{ville:1939};
more recent publications include
\cite{
dawid/vovk:1999,shafer/vovk:2001,\Takeuchi,kumon/etal:2007,horikoshi/takemura:2008,kumon/takemura:2008}).
The treatment of continuous time in \cite{shafer/vovk:2001} and \cite{GTP5}
uses non-standard analysis;
an important contribution of \cite{takeuchi/etal:2007}
is to avoid non-standard analysis
(which is both unfamiliar to many readers
and somewhat awkward in certain respects)
in studying the $\sqrt{\dd t}$ effect.
This paper also avoids non-standard analysis.
Its main result is Theorem \ref{thm:increase}
(all other results will be fairly obvious
to readers familiar with game-theoretic probability).

The words ``positive'', ``negative'',
``increasing'', ``decreasing'',
``before'', and ``after'' will be used in the wide sense
of the inequalities $\le$ or $\ge$, as appropriate;
we will add qualifiers ``strict'' or ``strictly''
when meaning the narrow sense of $<$ or $>$.
We will also be using the usual notation
$u\vee v:=\max(u,v)$, $u\wedge v:=\min(u,v)$,
and $u^+:=u\vee0$.

\ifarXiv
  The latest version of this working paper can be downloaded from the web site
  \texttt{http://probability%
  andfinance.com}
  (Working Paper 24).
\fi

\section{Null, almost certain, and completely uncertain events}
\label{sec:definitions}

The continuous time will be represented by the semi-infinite interval $[0,\infty)$.
We consider a perfect-information game between two players
called Reality and Sceptic.\footnote
  {Other names for these players,
  used in \cite{shafer/vovk:2001},
  are Market and Speculator, respectively.}
Reality outputs a continuous function $\omega:[0,\infty)\to\bbbr$,
interpreted as the price path of a financial asset
(although we do not insist on $\omega$ taking positive values),
and Sceptic tries to profit by trading in $\omega$.
First Sceptic presents his trading strategy and then Reality chooses $\omega$.
We start by formalizing this picture.

Let $\Omega$ be the set of all continuous functions $\omega:[0,\infty)\to\bbbr$.
For each $t\in[0,\infty)$,
$\FFF_t$ is defined to be the smallest $\sigma$-algebra
that makes all functions
$\omega\mapsto\omega(s)$, $s\in[0,t]$,
measurable.
A \emph{process} $S$ is a family of functions
$S_t:\Omega\to[-\infty,\infty]$, $t\in[0,\infty)$,
each $S_t$ being $\FFF_t$-measurable
(we drop ``stochastic'' since no probability measure on $\Omega$ is given,
and drop ``adapted'' for brevity).
An \emph{event} is an element of the $\sigma$-algebra
$\FFF_{\infty}:=\sigma(\cup_{t\in[0,\infty)}\FFF_t)$.
Stopping times $\tau:\Omega\to[0,\infty]$ w.r.\ to the filtration $(\FFF_t)$
and the corresponding $\sigma$-algebras $\FFF_{\tau}$
are defined as usual;
$\omega(\tau(\omega))$ and $S_{\tau(\omega)}(\omega)$
will be simplified to $\omega(\tau)$ and $S_{\tau}(\omega)$,
respectively.

The class of allowed strategies for Sceptic is defined in two steps.
An \emph{elementary trading strategy} $G$ consists of:
(a) an increasing infinite sequence of stopping times
$\tau_1\le\tau_2\le\cdots$ such that
$\lim_{n\to\infty}\tau_n(\omega)=\infty$ for each $\omega\in\Omega$;
(b) for each $n=1,2,\ldots$,
a bounded $\FFF_{\tau_{n}}$-measurable function $h_n$.
(It is possible that $\tau_n=\infty$ from some $n$ on,
which recovers the case of finite sequences.)
To such $G$ and an \emph{initial capital} $c\in\bbbr$
corresponds the \emph{elementary capital process}
\begin{equation}\label{eq:elementary-capital}
  \K^{G,c}_t(\omega)
  :=
  c
  +
  \sum_{n=1}^{\infty}
  h_n(\omega)
  \bigl(
    \omega(\tau_{n+1}\wedge t)-\omega(\tau_n\wedge t)
  \bigr),
  \quad
  t\in[0,\infty);
\end{equation}
the value $h_n(\omega)$ will be called the \emph{portfolio}
chosen at time $\tau_n$,
and $\K^{G,c}_t(\omega)$ will sometimes be referred to
as Sceptic's capital at time $t$.
Notice that the sum of finitely many elementary capital processes
is again an elementary capital process.

A \emph{positive capital process} is any process $S$
that can be represented in the form
\begin{equation}\label{eq:positive-capital}
  S_t(\omega)
  :=
  \sum_{n=1}^{\infty}
  \K^{G_n,c_n}_t(\omega),
\end{equation}
where the elementary capital processes $\K^{G_n,c_n}_t(\omega)$
are required to be positive, for all $t$ and $\omega$,
and the positive series $\sum_{n=1}^{\infty}c_n$ is required to converge
(intuitively,
the total capital invested has to be finite).
The sum (\ref{eq:positive-capital}) is always positive,
but we allow it to take value $+\infty$.
Since $\K^{G_n,c_n}_0(\omega)=c_n$ does not depend on $\omega$,
$S_0(\omega)$ also does not depend on $\omega$
and will sometimes be abbreviated to $S_0$.

The \emph{upper probability} of a set $E\subseteq\Omega$
is defined as
\begin{equation}\label{eq:upper-probability}
  \UpProb(E)
  :=
  \inf
  \bigl\{
    S_0
    \bigm|
    \forall\omega\in\Omega:
    \liminf_{t\to\infty}S_t(\omega)
    \ge
    \III_E(\omega)
  \bigr\},
\end{equation}
where $S$ ranges over the positive capital processes
and $\III_E$ stands for the indicator of $E$.
Notice that $\UpProb(\Omega)=1$
(in the terminology of \cite{shafer/vovk:2001},
our game protocol is ``coherent''):
indeed, $\UpProb(\Omega)<1$ would mean that some positive capital process
increases between time $0$ and $\infty$ for all $\omega\in\Omega$,
and this is clearly impossible for constant $\omega$.

We say that $E\subseteq\Omega$ is \emph{null} if $\UpProb(E)=0$,
and we say that $E$ is \emph{completely uncertain}
if $\UpProb(E)=1$ and $\UpProb(\Omega\setminus E)=1$.
A property of $\omega\in\Omega$ will be said to hold \emph{almost surely} (a.s.)\ if
the set of $\omega$ where it fails is null.
Correspondingly,
a set $E\subseteq\Omega$ is \emph{almost certain}
if $\UpProb(\Omega\setminus E)=0$.

\begin{remark}\label{rem:interpretation}
  \ifJOURNAL\begingroup\rm\fi
  The interpretation of almost certain events
  given in Section \ref{sec:introduction}
  was that we expect such events to happen in markets
  that are efficient to some degree;
  similarly, we do not expect null events to happen
  (provided such an event is singled out in advance).
  However, some qualifications are needed,
  since our definition of upper probability
  involves Sceptic's capital at infinity,
  which may be infinite without necessarily contradicting market efficiency.
  That interpretation is, e.g., valid for events $E\in\FFF_T$
  that happen or fail to happen before a finite horizon $T$:
  say, if $E$ is null,
  Sceptic can become arbitrarily rich by time $T$
  if $E$ happens.
  \ifJOURNAL\endgroup\fi
\end{remark}

The definition (\ref{eq:upper-probability}) enjoys a certain degree of robustness:
\begin{lemma}\label{lem:equivalence}
  We will obtain an equivalent definition
  replacing the $\liminf_{t\to\infty}$
  in (\ref{eq:upper-probability})
  by $\sup_{t\in[0,\infty)}$
  (and, therefore, by $\limsup_{t\to\infty}$).
\end{lemma}
\begin{proof}
  Suppose $\UpProb(E)<c<1$ in the sense of the definition with $\sup$
  and select a positive capital process witnessing this
  (i.e., satisfying $S_0<c$ and
  \begin{equation*}
    \forall\omega\in\Omega:
    \sup_{t\in[0,\infty)}S_t(\omega)
    \ge
    \III_E(\omega)
    ).
  \end{equation*}
  For any $\epsilon>0$,
  we can multiply $S$ by $1+\epsilon$ and stop it when it hits $1$;
  this will give a positive capital process
  witnessing $\UpProb(E)<(1+\epsilon)c$
  in the sense of the definition with $\liminf$.
  \ifJOURNAL\qed\fi
\end{proof}

Upper probability also enjoys the following useful property of $\sigma$-subadditivity
(obviously containing the property of finite subadditivity as a special case):
\begin{lemma}\label{lem:subadditivity}
  For any sequence of subsets $E_1,E_2,\ldots$ of $\Omega$,
  \begin{equation*}
    \UpProb
    \left(
      \bigcup_{n=1}^{\infty}
      E_n
    \right)
    \le
    \sum_{n=1}^{\infty}
    \UpProb(E_n).
  \end{equation*}
  In particular,
  a countable union of null sets is null.
\end{lemma}
\begin{proof}
  This follows immediately from the countability
  of a countable union of countable sets
  (of elementary capital processes).
  \ifJOURNAL\qed\fi
\end{proof}

The definition of a null set can be restated as follows.
\begin{lemma}
  A set $E\subseteq\Omega$ is null
  if and only if
  there exists a positive capital process $S$ with $S_0=1$
  such that $\lim_{t\to\infty}S_t(\omega)=\infty$ for all $\omega\in E$.
\end{lemma}
\begin{proof}
  Suppose $\UpProb(E)=0$.
  For each $n\in\{1,2,\ldots\}$,
  let $S^n$ be a positive capital process with $S^n_0=2^{-n}$
  and $\liminf_{t\to\infty}S^n_t\ge1$.
  It suffices to set $S:=\sum_{n=1}^{\infty}S^n$.
  \ifJOURNAL\qed\fi
\end{proof}

\section{Level sets of the price process}
\label{sec:level}

Our first theorem is a simple game-theoretic counterpart
of a standard measure-theoretic fact
(usually stated in the case of Brownian motion).
\begin{theorem}\label{thm:level}
  Let $b\in\bbbr$.
  Almost surely,
  the level set
  \begin{equation*}
    \LLL_{\omega}(b)
    :=
    \{t\in[0,\infty)\st\omega(t)=b\}
  \end{equation*}
  has no isolated points in $[0,\infty)$.
\end{theorem}
\begin{proof}
  If $\LLL_{\omega}(b)$ has an isolated point,
  there are rational numbers $a\ge0$ and $D\ne0$ such that
  strictly after the time $\inf\{t\st t\ge a,\omega(t)=b\}$
  $\omega$ does not take value $b$ before hitting the value $b+D$
  (this is true even if $0$ is the only isolated point of $\LLL_{\omega}(b)$).
  Suppose, for concreteness, that $D$ is positive
  (the case of negative $D$ is treated analogously).
  This event, which we denote $E_{a,D}$, is null:
  there is a positive capital process that starts from $\epsilon$
  (arbitrarily small positive number)
  and takes value $D+\epsilon$ when $E_{a,D}$ happens
  (choose portfolio $1$ at the time $\inf\{t\st t\ge a,\omega(t)=b\}$
  and then choose portfolio $0$ when the set $\{b-\epsilon,b+D\}$ is hit).
  Therefore,
  each event $E_{a,D}$ is null;
  it remains to apply Lemma \ref{lem:subadditivity}.
  \ifJOURNAL\qed\fi
\end{proof}

\begin{remark}
  \ifJOURNAL\begingroup\rm\fi
  As discussed in Remark \ref{rem:interpretation},
  almost certain events in $\FFF_T$ are expected to happen
  in markets that are efficient to some degree.
  The almost certain properties $E$ of sample paths
  that we establish in this paper do not belong to any $\FFF_T$, $T<\infty$,
  but it remains true that we expect them to happen
  in such markets.
  Each of these properties $E$ is ``falsifiable'' in the following sense:
  there exists a stopping time $\tau$,
  called a \emph{rejection time} for $E$,
  such that $E=\{\omega\st\tau(\omega)=\infty\}$.
  Moreover,
  it is possible to choose a rejection time $\tau$ for $E$
  such that for any monotonically increasing (however fast) function
  $f:[0,\infty)\to[0,\infty)$
  there exists a positive capital process $S$ with $S_0=1$
  such that $S_{\tau}(\omega)\ge f(\tau(\omega))$
  for all $\omega\in\Omega$ with $\tau(\omega)<\infty$.
  For example, the proof of Theorem \ref{thm:level} shows
  that Sceptic can become arbitrarily rich
  immediately after an isolated point in $\LLL_{\omega}(b)$ is observed.
  \ifJOURNAL\endgroup\fi
\end{remark}

\begin{corollary}\label{cor:level}
  For each $b\in\bbbr$,
  it is almost certain that the set $\LLL_{\omega}(b)$ is perfect,
  and so either is empty or has the cardinality of continuum.
\end{corollary}
\begin{proof}
  Since $\omega$ is continuous,
  the set $\LLL_{\omega}(b)$ is closed and so, by Theorem \ref{thm:level}, perfect.
  Non-empty perfect sets in $\bbbr$ always have the cardinality of continuum
  (see, e.g., \cite{\Aleksandrov}, Theorem 4.26).
  \ifJOURNAL\qed\fi
\end{proof}

The following lemma,
which uses some standard notions of measure-theoretic probability,
will allow us to show that many events of interest to us
are completely uncertain.
\begin{lemma}\label{lem:uncertain}
  Suppose $P(E)=1$,
  where $E$ is an event
  and $P$ is a probability measure on $(\Omega,\FFF_{\infty})$
  which makes the process $S_t(\omega):=\omega(t)$ a martingale
  w.r.\ to the filtration $(\FFF_t)$.
  Then $\UpProb(E)=1$.
\end{lemma}
\begin{proof}
  It suffices to prove that (\ref{eq:elementary-capital})
  is a local martingale under $P$.
  Indeed, in this case $\UpProb(E)<1$
  in conjunction with the maximal inequality for positive supermartingales
  would contradict the assumption that $P(E)=1$.
  It can be checked using the optional sampling theorem
  that each addend in (\ref{eq:elementary-capital}) is a martingale,
  and so each partial sum in (\ref{eq:elementary-capital}) is a martingale
  and (\ref{eq:elementary-capital}) itself is a local martingale.
  \ifSemiFULL

    In the rest of this proof I will check,
    for the sake of the readers with little experience in measure-theoretic probability
    (like myself),
    that each addend
    \begin{equation}\label{eq:addend}
      h_n(\omega)
      \bigl(
        \omega(\tau_{n+1}\wedge t)-\omega(\tau_n\wedge t)
      \bigr)
    \end{equation}
    in (\ref{eq:elementary-capital}) is indeed a martingale.
    For each $t\in[0,\infty)$,
    (\ref{eq:addend}) is integrable
    by the boundedness of $h_n$ and the optional sampling theorem
    (see, e.g., \cite{revuz/yor:1999}, Theorem II.3.2).
    We only need to prove, for $0<s<t$, that
    (omitting, until the end of the proof, the argument $\omega$ and ``a.s.'')
    \begin{equation}\label{eq:to-prove}
      \Expect
      \left(
        h_n
        \bigl(
          \omega(\tau_{n+1}\wedge t)-\omega(\tau_n\wedge t)
        \bigr)
        \mid
        \FFF_s
      \right)
      =
      h_n
      \bigl(
        \omega(\tau_{n+1}\wedge s)-\omega(\tau_n\wedge s)
      \bigr).
    \end{equation}
    We will check this equality on three $\FFF_s$-measurable events separately:
    \begin{description}
    \item[$\{\tau_{n+1}\le s\}$:]
      Both sides of the equality
      \begin{multline*}
        \Expect
        \left(
          h_n
          \bigl(
            \omega(\tau_{n+1}\wedge t)-\omega(\tau_n\wedge t)
          \bigr)
          \III_{\{\tau_{n+1}\le s\}}
          \mid
          \FFF_s
        \right)\\
        =
        h_n
        \bigl(
          \omega(\tau_{n+1}\wedge s)-\omega(\tau_n\wedge s)
        \bigr)
        \III_{\{\tau_{n+1}\le s\}}
      \end{multline*}
      are equal to the $\FFF_s$-measurable function
      $
        h_n
        \bigl(
          \omega(\tau_{n+1})-\omega(\tau_n)
        \bigr)
        \III_{\{\tau_{n+1}\le s\}}
      $
      (its $\FFF_s$-measurability follows, e.g.,
      from Lemma 1.2.15 in \cite{karatzas/shreve:1991} and the monotone-class theorem).
    \item[$\{\tau_n\le s<\tau_{n+1}\}$:]
      We need to check
      \begin{multline*}
        \Expect
        \left(
          h_n
          \bigl(
            \omega(\tau_{n+1}\wedge t)-\omega(\tau_n)
          \bigr)
          \III_{\{\tau_n\le s<\tau_{n+1}\}}
          \mid
          \FFF_s
        \right)\\
        =
        h_n
        \bigl(
          \omega(s)-\omega(\tau_n)
        \bigr)
        \III_{\{\tau_n\le s<\tau_{n+1}\}}.
      \end{multline*}
      Since $h_n\III_{\{\tau_n\le s<\tau_{n+1}\}}$ is bounded and $\FFF_s$-measurable,
      it suffices to check
      \begin{multline*}
        \Expect
        \left(
          \bigl(
            \omega(\tau_{n+1}\wedge t)-\omega(\tau_n)
          \bigr)
          \III_{\{\tau_n\le s<\tau_{n+1}\}}
          \mid
          \FFF_s
        \right)\\
        =
        \bigl(
          \omega(s)-\omega(\tau_n)
        \bigr)
        \III_{\{\tau_n\le s<\tau_{n+1}\}}.
      \end{multline*}
      Since $\omega(\tau_n)\III_{\{\tau_n\le s<\tau_{n+1}\}}$ is $\FFF_s$-measurable,
      it suffices to check
      \begin{equation*}
        \Expect
        \left(
          \omega(\tau_{n+1}\wedge t)
          \III_{\{\tau_n\le s<\tau_{n+1}\}}
          \mid
          \FFF_s
        \right)
        =
        \omega(s)
        \III_{\{\tau_n\le s<\tau_{n+1}\}},
      \end{equation*}
      which is the same thing as
      \begin{equation*}
        \Expect
        \left(
          \omega(s\vee\tau_{n+1}\wedge t)
          \III_{\{\tau_n\le s<\tau_{n+1}\}}
          \mid
          \FFF_s
        \right)
        =
        \omega(s)
        \III_{\{\tau_n\le s<\tau_{n+1}\}}
      \end{equation*}
      ($s\vee x\wedge t$ being a shorthand
      for $(s\vee x)\wedge t$ or, equivalently, $s\vee(x\wedge t)$).
      The stronger equality
      \begin{equation*}
        \Expect
        \left(
          \omega(s\vee\tau_{n+1}\wedge t)
          \mid
          \FFF_s
        \right)
        =
        \omega(s)
      \end{equation*}
      follows from the optional sampling theorem.
    \item[$\{s<\tau_n\}$:]
      We are required to prove
      \begin{equation*}
        \Expect
        \left(
          h_n
          \bigl(
            \omega(\tau_{n+1}\wedge t)-\omega(\tau_n\wedge t)
          \bigr)
          \III_{\{s<\tau_n\}}
          \mid
          \FFF_s
        \right)
        =
        0,
      \end{equation*}
      but we will prove more:
      \begin{equation*}
        \Expect
        \left(
          h_n
          \bigl(
            \omega(\tau_{n+1}\wedge t)-\omega(\tau_n\wedge t)
          \bigr)
          \III_{\{s<\tau_n\}}
          \mid
          \FFF_{s\vee\tau_n\wedge t}
        \right)
        =
        0.
      \end{equation*}
      Since the event $\{\tau_n\le t\}$,
      being equivalent to $\tau_n\le s\vee\tau_n\wedge t$,
      is $\FFF_{s\vee\tau_n\wedge t}$-measurable
      (see \cite{karatzas/shreve:1991}, Lemma~1.2.16),
      it is sufficient to prove
      \begin{equation}\label{eq:remains}
        \Expect
        \left(
          h_n
          \bigl(
            \omega(\tau_{n+1}\wedge t)-\omega(\tau_n\wedge t)
          \bigr)
          \III_{\{s<\tau_n\le t\}}
          \mid
          \FFF_{s\vee\tau_n\wedge t}
        \right)
        =
        0
      \end{equation}
      and
      \begin{equation*}
        \Expect
        \left(
          h_n
          \bigl(
            \omega(\tau_{n+1}\wedge t)-\omega(\tau_n\wedge t)
          \bigr)
          \III_{\{t<\tau_n\}}
          \mid
          \FFF_{s\vee\tau_n\wedge t}
        \right)
        =
        0.
      \end{equation*}
      The second equality is obvious,
      so our task has reduced to proving the first, (\ref{eq:remains}).
      Since $h_n\III_{\{\tau_n\le t\}}=h_n\III_{\{\tau_n\le s\vee\tau_n\wedge t\}}$
      is bounded and $\FFF_{s\vee\tau_n\wedge t}$-measurable,
      (\ref{eq:remains}) reduces to
      \begin{equation*}
        \Expect
        \left(
          \bigl(
            \omega(\tau_{n+1}\wedge t)-\omega(\tau_n\wedge t)
          \bigr)
          \III_{\{s<\tau_n\le t\}}
          \mid
          \FFF_{s\vee\tau_n\wedge t}
        \right)
        =
        0,
      \end{equation*}
      which is the same thing as
      \begin{equation*}
        \Expect
        \left(
          \bigl(
            \omega(s\vee\tau_{n+1}\wedge t)-\omega(s\vee\tau_n\wedge t)
          \bigr)
          \III_{\{s<\tau_n\le t\}}
          \mid
          \FFF_{s\vee\tau_n\wedge t}
        \right)
        =
        0.
      \end{equation*}
      The optional sampling theorem now gives
      \begin{multline*}
        \Expect
        \left(
          \bigl(
            \omega(s\vee\tau_{n+1}\wedge t)-\omega(s\vee\tau_n\wedge t)
          \bigr)
          \mid
          \FFF_{s\vee\tau_n\wedge t}
        \right)\\
        =
        \omega(s\vee\tau_n\wedge t) - \omega(s\vee\tau_n\wedge t)
        =
        0.
        \qedhere
      \end{multline*}
    \end{description}
    \fi
  \ifJOURNAL\qed\fi
\end{proof}

The following proposition shows that two standard properties
of typical sample paths of Brownian motion
become completely uncertain for continuous price processes.
\begin{proposition}\label{prop:level}
  Let $b\in\bbbr$.
  The following events are completely uncertain:
  \begin{enumerate}
  \item\label{it:level-Lebesgue}
    the Lebesgue measure of $\LLL_{\omega}(b)$ is zero;
  \item\label{it:level-bounded}
    the set $\LLL_{\omega}(b)$ is unbounded.
  \end{enumerate}
\end{proposition}
\begin{proof}
  To see that the upper probability
  of \ref{it:level-bounded} and of the complement of \ref{it:level-Lebesgue}
  is $1$,
  consider the martingale
  that is identically equal to $b$.
  To see that the upper probability
  of \ref{it:level-Lebesgue} and of the complement of \ref{it:level-bounded}
  is $1$,
  consider a constant martingale not equal to $b$.
  (Notice that these arguments do not really require Lemma \ref{lem:uncertain}.)
  \ifJOURNAL\qed\fi
\end{proof}

\section{Properties related to non-increase}
\label{sec:increase}

Let us say that $t\in[0,\infty)$ is a \emph{point of semi-strict increase} for $\omega$
if there exists $\delta>0$
such that $\omega(s)\le\omega(t)<\omega(u)$
for all $s\in((t-\delta)^+,t)$ and $u\in(t,t+\delta)$.
Points of semi-strict decrease are defined in the same way
except that $\omega(s)\le\omega(t)<\omega(u)$
is replaced by $\omega(s)\ge\omega(t)>\omega(u)$.
The following theorem is the game-theoretic counterpart
of Dvoretzky, Erd\H{o}s, and Kakutani's \cite{dvoretzky/etal:1961}
result for Brownian motion
(Dubins and Schwarz \cite{dubins/schwarz:1965}
noticed that it continues to hold for all continuous martingales);
its proof (based on Krzysztof Burdzy's idea) can be found in Appendix B.
\begin{theorem}\label{thm:increase}
  Almost surely,
  $\omega$ has no points of semi-strict increase or decrease.
\end{theorem}

We will also state several corollaries of Theorem \ref{thm:increase}.
First, the price process is nowhere monotone (unless constant):
\begin{corollary}\label{cor:monotone}
  Almost surely,
  $\omega$ is monotone in no open interval,
  unless it is constant in that interval.
\end{corollary}
\begin{proof}
  This is an obvious corollary of Theorem \ref{thm:increase},
  but is also easy to prove directly:
  each interval of monotonicity where $\omega$ is not constant
  contains a rational time point $a$
  after which $\omega$ increases
  (if we assume, for concreteness,
  that ``monotonicity'' means ``increase'')
  by a rational amount $D>0$
  before hitting the level $\omega(a)$ again;
  as in the proof of Theorem \ref{thm:level},
  it is easy to show that this event,
  denoted $E_{a,D}$,
  is null,
  and it remains to apply Lemma \ref{lem:subadditivity}
  to deduce that $\cup_{a,D}E_{a,D}$ is also null.
  \ifJOURNAL\qed\fi
\end{proof}

Let us say that a closed interval $[t_1,t_2]\subseteq[0,\infty)$
is an \emph{interval of local maximum} for $\omega$
if (a) $\omega$ is constant on $[t_1,t_2]$
but not constant on any larger interval containing $[t_1,t_2]$,
and (b) there exists $\delta>0$ such that $\omega(s)\le\omega(t)$
for all $s\in((t_1-\delta)^+,t_1)\cup(t_2,t_2+\delta)$
and all $t\in[t_1,t_2]$.
In the case where $t_1=t_2$
we can say ``point'' instead of ``interval''.
A ray $[t,\infty)$, $t\in[0,\infty)$,
is a \emph{ray of local maximum} for $\omega$
if (a) $\omega$ is constant on $[t,\infty)$
but not constant on any larger ray $[s,\infty)$, $s\in(0,t)$,
and (b) there exists $\delta>0$ such that $\omega(s)\le\omega(t)$
for all $s\in((t-\delta)^+,t)$.
An \emph{interval} or \emph{ray of strict local maximum}
is defined in the same way
except that $\omega(s)\le\omega(t)$ is replaced by $\omega(s)<\omega(t)$.
The definitions of intervals and rays of (strict) local minimum
are obtained by obvious modifications;
as usual ``extremum'' means maximum or minimum.
We say that $t\in[0,\infty)$ is a \emph{point of constancy} for $\omega$
if there exists $\delta>0$
such that $\omega(s)=\omega(t)$ for all $s\in((t-\delta)^+,t+\delta)$;
points $t\in[0,\infty)$ that are not points of constancy are \emph{points of non-constancy}.
(Notice that we do not count points of constancy among points of local extremum.)
\begin{corollary}\label{cor:maxima}
  Almost surely,
  every interval of local extremum is a point,
  all points and the ray (if it exists) of local extremum are strict,
  the set of points of local extremum is countable,
  and any neighbourhood of any point of non-constancy
  contains a point of local maximum and a point of local minimum.
\end{corollary}
\begin{proof}
  We will prove only the statements concerning local maxima.

  If $\omega$ has an interval of local maximum $[t_1,t_2]$
  with $t_1\ne t_2$,
  $t_2$ will be a point of semi-strict decrease,
  and by Theorem \ref{thm:increase}
  it is almost certain that there will be no such points
  (alternatively, one could use the direct argument
  given in the proof of Corollary~\ref{cor:monotone}).
  We can see that no such $[t_1,t_2]$
  can even be an interval of local maximum ``on the right''.

  Now suppose that there is a point or ray of local maximum that is not strict.
  In this case there is a quadruple $0<t_1<t_2<t_3<t_4$
  of rational numbers and another rational number $D>0$ such that
  $\max_{t\in[t_1,t_2]}\omega(t)=\max_{t\in[t_3,t_4]}\omega(t)>\omega(t_4)+D$.
  The event that such a set of rational numbers exists is null:
  proceed as in the proof of Theorem \ref{thm:level}.

  The set of all points of strict local maximum is countable,
  as the following standard argument
  demonstrates:
  each point of strict local maximum can be surrounded
  by an open interval with rational end-points
  in which that point is a strict maximum,
  and all these open intervals will be different.

  Finally,
  Corollary \ref{cor:monotone} immediately implies
  that every neighbourhood of every point of non-constancy
  contains a point of local maximum.
  \ifJOURNAL\qed\fi
\end{proof}

This is a simple game-theoretic version
of the classical result about nowhere differentiability of Brownian motion
(Paley, Wiener, and Zygmund 
\cite{paley/etal:1933}):
\begin{corollary}\label{cor:derivative}
  Almost surely,
  $\omega$ does not have a non-zero derivative anywhere.
\end{corollary}
\begin{proof}
  A point where a non-zero derivative exists is a point of semi-strict increase or decrease.
  \ifJOURNAL\qed\fi
\end{proof}
\noindent
It would interesting to find stronger versions of the Paley--Wiener--Zygmund result
(although see parts \ref{it:const} and \ref{it:no} of Proposition \ref{prop:increase}).

The following proposition demonstrates
the necessity of various conditions
in Corollaries \ref{cor:monotone}--\ref{cor:derivative}.
\begin{proposition}\label{prop:increase}
  The following events are completely uncertain:
  \begin{enumerate}
  \item\label{it:const}
    $\omega$ is constant on $[0,\infty)$;
  \item\label{it:extremum}
    for some $t\in(0,\infty)$,
    $[t,\infty)$ is the ray of local maximum (or minimum) for~$\omega$;
  \item\label{it:no}
    $\omega'(t)$ exists for no $t\in[0,\infty)$.
  \end{enumerate}
\end{proposition}
\begin{proof}
  We will be using Lemma \ref{lem:uncertain}.
  To see that the upper probability of \ref{it:no},
  of the complement of \ref{it:const}, and of the complement of \ref{it:extremum} is $1$,
  remember that Brownian motion is a martingale.
  To see that the upper probability of \ref{it:const} and of the complement of \ref{it:no} is $1$,
  consider a constant martingale.
  To see that the upper probability of \ref{it:extremum} is $1$,
  consider the following continuous martingale:
  start as Brownian motion from $0$ and stop when $1$ (or $-1$) is hit.
  \ifJOURNAL\qed\fi
\end{proof}

\section{Conclusion}

This paper gives provisional definitions
of upper probability and related notions
(such as that of null events)
for the case of continuous time.
It might stay too close to the standard measure-theoretic framework
in that the flow of information is modelled
as a filtration.
In discrete-time game-theoretic probability,
as presented in \cite{shafer/vovk:2001},
measurability does not play any special role,
whereas in measure-theoretic probability
measurability has the obvious technical role to play.
On one hand,
we could drop all conditions of measurability in all the definitions given above
(equivalently, replace each $\sigma$-algebra that we used
by the smallest class of subsets of $\Omega$
containing that $\sigma$-algebra and closed
under arbitrary unions and intersections);
it is obvious that all our theorems and corollaries,
Proposition \ref{prop:level}, and parts of Proposition \ref{prop:increase}
still hold
(and it is an interesting problem to establish
whether the remaining parts of Proposition \ref{prop:increase} continue to hold).
On the other hand,
one might want to strengthen the requirement of measurability
to that of computability.

\subsection*{Acknowledgments}

This work was partially supported by EPSRC (grant EP/F002998/1),
MRC (grant G0301107),
and the Cyprus Research Promotion Foundation.


\appendix
\ifJOURNAL
  \section{A one-sided law of large numbers}
\fi
\ifnotJOURNAL
  \section*{Appendix A: A one-sided law of large numbers}
\fi

In this appendix we establish a result
that will be needed in the proof of Theorem \ref{thm:increase}.
This result involves the following perfect-information game protocol
depending on two parameters, $N\in\{1,2,\ldots\}$ (the horizon) and $c>0$:

\medskip

\noindent
\textbf{Players:} Reality, Sceptic

\parshape=6
\IndentI  \WidthI
\IndentI  \WidthI
\IndentII \WidthII
\IndentII \WidthII
\IndentII \WidthII
\IndentI  \WidthI
\noindent
$\K_0 := 1$.\\
FOR $n=1,2,\dots,N$:\\
  Sceptic announces $s_n\ge0$.\\
  Reality announces $x_n\in[-c,c]$.\\
  $\K_{n} := \K_{n-1} + s_n x_n$.\\
END FOR

\medskip

\noindent
This is a one-sided version of the fair-coin game in \cite{shafer/vovk:2001}, p.~124
(intuitively,
the restriction $s_n\ge0$ means
that the expected value of $x_n$ is zero or strictly negative,
and $\K_n$ is interpreted as Sceptic's capital).
A strategy for Sceptic is \emph{prudent}
if it guarantees $\K_n\ge0$, for all $n$ and regardless of Reality's moves.
The definition of upper probability in this simple discrete-time case becomes
\begin{multline*}
  \UpProb(E)
  :=
  \inf
  \bigl\{
    \delta
    \bigm|
    \text{Sceptic has a prudent strategy}\\
    \text{that guarantees $\K_N\ge1/\delta$ when $(x_1,\ldots,x_N)\in E$}
  \bigr\},
\end{multline*}
where $E$ is a subset of the \emph{sample space} $[-c,c]^N$.

The following lemma is a simple game-theoretic one-sided weak law of large numbers.
\begin{lemma}\label{lem:WLLN}
  Let $\delta_1 > 0$, $\delta_2 > 0$, and $N \ge c^2/\delta_1 \delta_2^2$.
  Then
  \begin{equation*}
    \UpProb
    \left(
       \frac1N
       \sum_{n=1}^N
       x_n
       \ge
       \delta_2
    \right)
    \le
    \delta_1.
  \end{equation*}
\end{lemma}

In the proof of Theorem \ref{thm:increase}
we will actually need the following more precise version of Lemma \ref{lem:WLLN}:
\begin{lemma}\label{lem:WLLN-capital}
  Sceptic has a strategy that guarantees that his capital $\K_n$ will satisfy
  \begin{equation}\label{eq:capital1}
    \K_n
    \ge
    \frac{N-n}{N}
    +
    \frac{1}{c^2N}
    \left(
      \sum_{j=1}^n
      x_j
    \right)^{+,2}
  \end{equation}
  for $n=0,1,\dots,N$, where $t^{+,2}:=(t^+)^2$.
\end{lemma}
\begin{proof}
  This proof is based on the idea used in \cite{shafer/vovk:arXiv0706.3188}
  (proof of Lemma~2).
  When $n=0$, (\ref{eq:capital1}) reduces to $\K_0 \ge 1$,
  which we know is true.
  So it suffices to show that if (\ref{eq:capital1}) holds for $n<N$,
  then Sceptic can make sure that the corresponding inequality for $n+1$,
  \begin{equation}\label{eq:capital2}
    \K_{n+1}
    \ge
    \frac{N-n-1}{N} +
    \frac{1}{c^2N}
    \left(
      \sum_{j=1}^{n+1}
      x_j
    \right)^{+,2},
\end{equation}
also holds.
This is how Sceptic chooses his move:
\begin{itemize}
\item
  If
  $
    \sum_{j=1}^n
    x_j
    \ge
    0
  $,
  then Sceptic sets
  \begin{equation}\label{eq:M}
    s_{n+1}
    :=
    \frac{2}{c^2N}
    \sum_{j=1}^n
    x_j
    \ge
    0.
  \end{equation}
  In this case
  \begin{align*}
    \K_{n+1}
    &=
    \K_n
    +
    \frac{2}{c^2N}
    \left(
      \sum_{j=1}^n
      x_j
    \right)
    x_{n+1}\notag\\
    &\ge
    \frac{N-n}{N}
    +
    \frac{1}{c^2N}
    \left(
      \sum_{j=1}^n
      x_j
    \right)^2
    +
    \frac{2}{c^2N}
    \left(
      \sum_{j=1}^n
      x_j
    \right)
    x_{n+1}\notag\\
    &=
    \frac{N-n}{N}
    +
    \frac{1}{c^2N}
    \left(
      \sum_{j=1}^{n+1}
      x_j
    \right)^2
    -
    \frac{x_{n+1}^2}{c^2N}\notag\\
    &\ge
    \frac{N-n-1}{N}
    +
    \frac{1}{c^2N}
    \left(
      \sum_{j=1}^{n+1}
      x_j
    \right)^{+,2} 
  \end{align*}
  (the last inequality uses $t^2\ge t^{+,2}$),
  which coincides with (\ref{eq:capital2}).
\item
  If
  $
    \sum_{j=1}^n
    x_j
    <
    0
  $,
  Sceptic sets $s_{n+1}:=0$, and so $\K_{n+1}=\K_n$.
  Because
  \begin{equation*}
    \left(
      \sum_{j=1}^{n+1}
      x_j
    \right)^{+,2}
    -
    \left(
      \sum_{j=1}^n
      x_j
    \right)^{+,2}
    \le
    x_{n+1}^2
    \le
    c^2,
  \end{equation*}
  we again obtain~(\ref{eq:capital2}) from~(\ref{eq:capital1}).
  \ifJOURNAL\qed\fi
\end{itemize}
\end{proof}

\ifJOURNAL
  \begin{proof*}{Proof of Lemma~\ref{lem:WLLN}.}
\fi
\ifnotJOURNAL
  \begin{proof}[Proof of Lemma~\ref{lem:WLLN}]
\fi
  Sceptic's strategy in Lemma~\ref{lem:WLLN-capital} is prudent
  (it is obvious that $\K_n \ge 0$ for all $n$),
  and~(\ref{eq:capital1}) implies
  \begin{equation*}
    \K_N
    \ge
    \frac{1}{c^2N}
    \left(
      \sum_{n=1}^{N}
      x_n
    \right)^{+,2}.
  \end{equation*}
  Combining this inequality with the assumption that
  $N \ge c^2/\delta_1\delta_2^2$,
  we see that when the event $\frac1N\sum_{n=1}^Nx_n \ge \delta_2$ happens,
  $\K_N \ge 1/\delta_1$.
  \ifJOURNAL\qed\fi
\ifJOURNAL
  \end{proof*}
\fi
\ifnotJOURNAL
  \end{proof}
\fi

\ifJOURNAL
  \section{Proof of Theorem \ref{thm:increase}}
\fi
\ifnotJOURNAL
  \section*{Appendix B: Proof of Theorem \ref{thm:increase}}
\fi

This proof is modelled on the very simple proof
of Dvoretzky, Erd\H{o}s, and Kakutani's result
given by Burdzy \cite{burdzy:1990}.
We will only prove that, almost surely,
$\omega$ has no points of semi-strict increase in $(0,\infty)$
(the argument given in the direct proof of Corollary \ref{cor:monotone},
with $a=0$,
shows that almost surely $0$ cannot be a point of semi-strict increase).

It suffices to prove that,
for any given positive constants $C$ and $D$,
the following event, denoted $E_{C,D}$, is null:
the price process $\omega$ starts from 0,
reaches a point of semi-strict increase $t$ before hitting the level $C$,
then reaches the level $\omega(t)+D$ before hitting $\omega(t)$ again.
Indeed, suppose $\omega$ in the original game has a point of semi-strict increase,
say $t>0$.
There are positive rational numbers $a\in[0,t)$, $C$, and $D$
such that $\omega(s)\le\omega(t)\le\omega(a)+C$, for all $s\in[a,t)$,
and $\omega$ hits $\omega(t)+D$ before hitting $\omega(t)$ strictly after moment $t$.
The latter event, denoted by $E_{a,C,D}$ is null
since it is a translation of the null event $E_{C,D}$.
By Lemma \ref{lem:subadditivity},
the union of all $E_{a,C,D}$ is also null,
which completes the proof.

Fix positive $C$ and $D$;
our goal is to prove that $E_{C,D}$ is null.
For each $\epsilon\in(0,1)$ (intuitively, a small constant),
define sequences of stopping times $U_n$ and $T_n$
and a sequence of functions $M_n$ on $\Omega$,
$n=0,1,\ldots$, as
\begin{align*}
  &M_0 := 0,
  \quad
  U_0 := 0,\\
  &T_n
  :=
  \inf\bigl\{t>U_n\bigm|\omega(t)\in\{M_n-\epsilon,M_n+D\}\bigr\},
  \quad
  n=0,1,\ldots,\\
  &M_{n+1}
  :=
  \sup\bigl\{\omega(t)\bigm|t\in[0,T_n)\bigr\},
  \quad
  n=0,1,\ldots,\\
  &U_{n+1}
  :=
  \inf\bigl\{t>T_n\bigm|\omega(t)=M_{n+1}\bigr\},
  \quad
  n=0,1,\ldots;
\end{align*}
as usual, $\inf\emptyset$ is interpreted as $\infty$.
We also set
\begin{equation*}
  X_n := M_n - M_{n-1}, \enspace n=1,2,\ldots,
  \quad
  N
  :=
  \left\lfloor
    \frac{1}{\epsilon\sqrt{\ln\frac{1}{\epsilon}}}
  \right\rfloor.
\end{equation*}
It suffices to establish,
for an arbitrarily large constant $K>0$,
the existence of two positive elementary capital processes
having strictly positive initial values
and satisfying the following conditions when $\omega\in E_{C,D}$:
\begin{enumerate}
\item\label{it:first}
  The first process increases $K$-fold
  if $T_{N-1}<\infty$
  and the price level $C$ is not attained before time $T_{N-1}$.
\item\label{it:second}
  The second process increases $K$-fold
  if $T_{N-1}=\infty$
  or the price level $C$ is attained before time $T_{N-1}$.
\end{enumerate}
An elementary trading strategy
leading to \ref{it:second}
chooses portfolio $1$ at time $U_0$, portfolio $0$ at time $T_0$,
portfolio $1$ at time $U_1$, portfolio $0$ at time $T_1$, etc.;
finally, portfolio $1$ at time $U_{N-1}$ and portfolio $0$ at time $T_{N-1}$.
The strategy is started with initial capital $\epsilon N$
to ensure that its capital process is positive.
If $T_{N-1}=\infty$
or the price level $C$ is attained before $T_{N-1}$,
it will be true that $\omega(T_n)=M_n+D$ for some $n\in\{0,\ldots,N-1\}$,
and so the final capital will be at least $D$.
By the definition of $N$,
we can ensure $\epsilon N\le D/K$ by choosing a small $\epsilon$.

It remains to prove the existence of an elementary trading strategy
leading to \ref{it:first}.
Intuitively, this strategy will implement a law of large numbers;
in this paragraph we will discuss the situation informally
considering the case of Brownian motion.
Why can we expect that the price level $C$ will be attained?
For $x\ge0$,
\begin{equation*}
  \Prob\{X_n\ge x\}
  =
  \begin{cases}
    \epsilon/(x+\epsilon) & \text{if $x\in[0,D]$}\\
    0 & \text{otherwise},
  \end{cases}
\end{equation*}
and so we can compute the expectation of the truncated version
$\tilde X_n := X_n \wedge (\sqrt{\epsilon}-\epsilon)$
of $X_n$ as
\begin{equation*}
  \Expect\tilde X_n
  =
  \int_0^{\sqrt{\epsilon}-\epsilon}
  \frac{\epsilon\dd x}{x+\epsilon}
  =
  \frac{\epsilon}{2}\ln\frac{1}{\epsilon}
\end{equation*}
($\epsilon$ is assumed small throughout;
in particular, $\sqrt{\epsilon}-\epsilon\le D$);
it is clear that the variance of $\tilde X_n$ does not exceed $\epsilon$.
The expectation of the sum
$\tilde X_1 + \cdots + \tilde X_N \le X_1+\cdots+X_N$
will exceed or be approximately equal to
$N\frac{\epsilon}{2}\ln\frac{1}{\epsilon}\approx\frac12\sqrt{\ln\frac{1}{\epsilon}}\gg1$
and its variance will be at most $N\epsilon\approx 1/\sqrt{\ln\frac{1}{\epsilon}}\ll1$.
Therefore, the sum of $X_n$ can be expected to exceed $C$.
The purpose of this paragraph has been to get a sense of direction
in which we are moving,
and now we resume the actual proof.

Choose a positive constant $\delta>0$
(intuitively, small even as compared with $\epsilon$)
such that the ratio $M:=(\sqrt{\epsilon}-\epsilon)/\delta$ is integer.
The Darboux sums for the Riemann integral used earlier
for computing $\Expect\tilde X_n$ are
\begin{equation*}
  L
  :=
  \sum_{m=1}^M
  \frac{\epsilon\delta}{m\delta+\epsilon}
  \le
  \int_0^{\sqrt{\epsilon}-\epsilon}
  \frac{\epsilon\dd x}{x+\epsilon}
  \le
  \sum_{m=0}^{M-1}
  \frac{\epsilon\delta}{m\delta+\epsilon};
\end{equation*}
we will be interested in the lower Darboux sum $L$.
Fix temporarily an $n\in\{1,\ldots,N\}$.
For each $m\in\{1,\ldots,M\}$,
there is a positive elementary capital process
starting at time $U_{n-1}$ from $\epsilon\delta/(m\delta+\epsilon)$
and ending at:
\begin{itemize}
\item
  $\delta$ if and when $\omega$ hits $M_{n-1}+m\delta$
  (provided this happens before $T_{n-1}$);
\item
  $0$ if and when $\omega$ hits $M_{n-1}-\epsilon$ (at time $T_{n-1}$)
  before hitting $M_{n-1}+m\delta$.
\end{itemize}
Indeed, such a process can be obtained
by choosing portfolio $\delta/(m\delta+\epsilon)$ at time $U_{n-1}$
and then choosing portfolio 0
when $M_{n-1}+m\delta$ or $M_{n-1}-\epsilon$ is hit.
The sum $S_n$ of such elementary capital processes over $m=1,\ldots,M$
will also be a positive elementary capital process.

The initial capital $S_n(U_{n-1})$ of $S_n$ is $L$,
and it is easy to see that
$S_n(T_{n-1})=\delta\lfloor\tilde X_n/\delta\rfloor\le\tilde X_n$.
The elementary capital process $L-S_n$ starts from $0$
and ends up with at least $x_n:=L-\tilde X_n$ at time $T_{n-1}$.

Let us take $\delta$ so small that
$L\ge\frac{\epsilon}{3}\ln\frac{1}{\epsilon}$.
Lemma \ref{lem:WLLN-capital} gives an explicit elementary capital process $\K$
that starts from $1$ and ends with at least
\begin{multline}\label{eq:final}
  \frac
  {1}
  {
    \sqrt{\epsilon}^2
    \left\lfloor
      \frac{1}{\epsilon\sqrt{\ln\frac{1}{\epsilon}}}
    \right\rfloor
  }
  \left(
    \frac{\epsilon}{3}\ln\frac{1}{\epsilon}
    \left\lfloor
      \frac{1}{\epsilon\sqrt{\ln\frac{1}{\epsilon}}}
    \right\rfloor
    -
    \sum_{n=1}^N
    \tilde X_n
  \right)^{+,2}\\
  \ge
  \sqrt{\ln\frac{1}{\epsilon}}
  \left(
    \frac{1}{4}\sqrt{\ln\frac{1}{\epsilon}}
    -
    \sum_{n=1}^N
    \tilde X_n
  \right)^{+,2}
\end{multline}
at time $T_{N-1}$.
On the event $\sum_{n=1}^N\tilde X_n\le C$,
the final capital (\ref{eq:final}) can be made arbitrarily large
by choosing a small $\epsilon$.

We still need to make sure that the elementary capital process $\K$
constructed in the last paragraph is positive:
we did not show that it does not become strictly negative
strictly between $U_{n-1}$ and $T_{n-1}$.
According to (\ref{eq:M}),
Sceptic's move $s_n$ never exceeds
\begin{equation*}
  \frac
  {2}
  {
    \sqrt{\epsilon}^2
    N
  }
  N
  \frac{\epsilon}{2}
  \ln\frac{1}{\epsilon}
  =
  \ln\frac{1}{\epsilon},
\end{equation*}
and
\begin{equation*}
  S_n
  \le
  \sum_{m=1}^M
  \frac{\delta}{m\delta+\epsilon}
  m\delta
  \le
  M\delta
  \le
  \sqrt{\epsilon}
\end{equation*}
implies $L-S_n\ge-\sqrt{\epsilon}$.
Therefore,
our elementary capital process $\K$
is always at least $-\sqrt{\epsilon}\ln\frac{1}{\epsilon}$;
$\sqrt{\epsilon}\ln\frac{1}{\epsilon}$
is a small amount that can be added to the initial capital
to make $\K$ positive.
This completes the proof.


\begin{thebibliography}{10}

\bibitem{aleksandrov:1977latin}
Pavel~S\DOT{} Aleksandrov.
\newblock {\em Vvedenie v teoriyu mnozhestv i obshchuyu to\-po\-lo\-gi\-yu
  (Introduction to Set Theory and General Topology, in Russian)}.
\newblock Nauka, Moscow, 1977.

\bibitem{burdzy:1990}
Krzysztof Burdzy.
\newblock On nonincrease of {B}rownian motion.
\newblock {\em Annals of Probability}, 18:978--980, 1990.

\bibitem{dawid/vovk:1999}
A\DOT{}~Philip Dawid and \Vladimir{} Vovk.
\newblock Prequential probability: principles and properties.
\newblock {\em Bernoulli}, 5:125--162, 1999.

\bibitem{dubins/schwarz:1965}
Lester~E\DOT{} Dubins and Gideon Schwarz.
\newblock On continuous martingales.
\newblock {\em Proceedings of the National Academy of Sciences}, 53:913--916,
  1965.

\bibitem{dvoretzky/etal:1961}
Aryeh Dvoretzky, Paul Erd\H{o}s, and Shizuo Kakutani.
\newblock Nonincrease everywhere of the {B}rownian motion process.
\newblock In {\em Proceedings of the Fourth Berkeley Symposium on Mathematical
  Statistics and Probability}, volume~2, pages 103--116, Berkeley, CA, 1961.
  University of California Press.

\bibitem{horikoshi/takemura:2008}
Yasunori Horikoshi and Akimichi Takemura.
\newblock Implications of contrarian and one-sided strategies for the fair-coin
  game.
\newblock \emph{Stochastic Processes and their Applications}, to appear,
  doi:10.1016/j.spa.2007.11.007.

\bibitem{karatzas/shreve:1991}
Ioannis Karatzas and Steven~E\DOT{} Shreve.
\newblock {\em Brownian Motion and Stochastic Calculus}.
\newblock Springer, New York, second edition, 1991.

\bibitem{kumon/takemura:2008}
Masayuki Kumon and Akimichi Takemura.
\newblock On a simple strategy weakly forcing the strong law of large numbers
  in the bounded forecasting game.
\newblock \emph{Annals of the Institute of Statistical Mathematics}, to appear.

\bibitem{kumon/etal:2007}
Masayuki Kumon, Akimichi Takemura, and Kei Takeuchi.
\newblock Game-theoretic versions of strong law of large numbers for unbounded
  variables.
\newblock {\em Stochastics}, 79:449--468, 2007.

\bibitem{paley/etal:1933}
Raymond E\DOT~A\DOT~C\DOT{} Paley, Norbert Wiener, and Antoni Zygmund.
\newblock Note on random functions.
\newblock {\em Mathematische Zeitschrift}, 37:647--668, 1933.

\bibitem{shafer/vovk:2001}
Glenn Shafer and \Vladimir{} Vovk.
\newblock {\em Probability and Finance: It's Only a Game!}
\newblock Wiley, New York, 2001.

\bibitem{shafer/vovk:arXiv0706.3188}
Glenn Shafer and \Vladimir{} Vovk.
\newblock A tutorial on conformal prediction.
\newblock Technical Report \texttt{arXiv:cs/0706.3188} [cs.LG],
  \texttt{arXiv.org} e-Print archive, June 2007.
\newblock To appear in \emph{Journal of Machine Learning Research}.

\bibitem{takeuchi:2004latin}
Kei Takeuchi.
\newblock {\em Kake no suuri to kinyu kogaku (Mathematics of Betting and
  Financial Engineering, in Japanese)}.
\newblock Saiensusha, Tokyo, 2004.

\bibitem{takeuchi/etal:2007}
Kei Takeuchi, Masayuki Kumon, and Akimichi Takemura.
\newblock A new formulation of asset trading games in continuous time with
  essential forcing of variation exponent.
\newblock Technical Report \texttt{arXiv:0708.0275v1} [math.PR],
  \texttt{arXiv.org} e-Print archive, August 2007.

\bibitem{ville:1939}
Jean Ville.
\newblock {\em Etude critique de la notion de collectif}.
\newblock Gauthier-Villars, Paris, 1939.

\bibitem{GTP5}
\Vladimir{} Vovk and Glenn Shafer.
\newblock A game-theoretic explanation of the $\sqrt{dt}$ effect. {T}he
  {G}ame-{T}heoretic {P}robability and {F}inance project,
  \texttt{http://prob\linebreak[0]a\linebreak[0]bil\linebreak[0]i\linebreak[0]%
ty\linebreak[0]and\linebreak[0]fi\linebreak[0]nance.com}, {W}orking {P}aper 5,
  January 2003.

\end{thebibliography}
\end{document}